\documentclass[12pt]{article}
\usepackage{pdproc,epsfig} 

\begin{document}

\renewcommand{\thefootnote}{\alph{footnote}}
  
\title{
 IceCube: Status and Results}

\author{ Thomas K. Gaisser}

\address{ Bartol Research Institute and Dept. of Physics \& Astronomy\\
University of Delaware\\
Newark, DE 19716, USA\\
 {\rm E-mail: gaisser@bartol.udel.edu}}

  \centerline{\footnotesize}

\author{For the IceCube Collaboration\\
{\rm url: http://www.icecube.wisc.edu/collaboration/}}

\abstract{This talk describes the complete IceCube neutrino telescope and
summarizes some results obtained while the detector was under construction.}
   
\normalsize\baselineskip=15pt

\section{Introduction}
\begin{figure}
     \mbox{\epsfig{figure=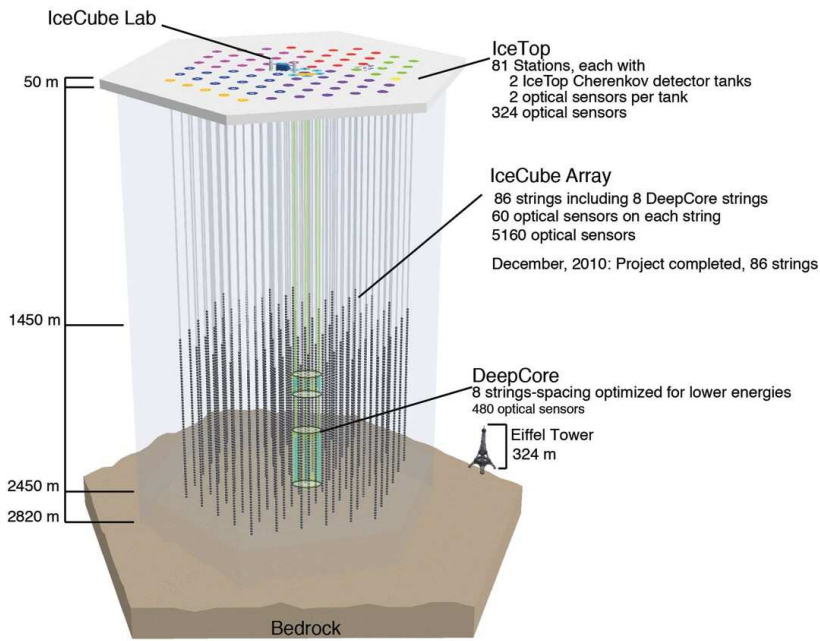,width=12.5cm}}
\caption{Artist's drawing of IceCube.}
\label{detector}
\end{figure}

  On December 18, 2010 the last cable of IceCube was lowered into position, thus completing
  construction of the first kilometer-scale neutrino detector.  After commissioning
  of the optical modules on seven newly deployed strings, the full IceCube detector with 86 strings
  and 81 IceTop stations on the surface was turned on May 20, 2011.  There are 5160
  optical modules between 1450 and 2450 meters in the deep
  ice and 324 optical modules in 81 pairs of tanks on the surface, as illustrated
  in Fig.~\ref{detector}.
  
  IceCube collected data in various configurations during construction.  In this paper 
  results obtained from April 2008 into May 2009 are presented, along with
  some more recent data to illustrate performance of the detector.  In 2008/09 
  IceCube was half complete with 40 strings viewing half a cubic kilometer of deep ice and an air
  shower array with 40 stations on the surface above.  To set the context for the results, the
  paper begins with a historical introduction followed by an account of the design, construction
  and operation of IceCube.
  
  \section{Historical background}
  The idea of instrumenting a large volume of water as a target for naturally occurring
  neutrinos was discussed fifty years ago by Reines~\cite{Reines}, by Greisen~\cite{Greisen}
  and by Markov~\cite{Markov}.  Cherenkov radiation from charged particles produced in the
  interactions of neutrinos would propagate over many meters and allow reconstruction
  of events with relatively few optical modules.  Markov described how the neutrinos
  produced by interactions of cosmic rays in the atmosphere might be used to
  study the energy dependence of the neutrino cross section.  Greisen wanted 
  to search for high energy neutrinos of astrophysical origin above the
  background of the steeper atmospheric neutrino spectrum.  Reines briefly noted the likely existence
  of cosmic neutrinos produced by interactions of cosmic rays in extraterrestrial sources.  He then
  went on to estimate the interaction
  rate of cosmic-ray neutrinos produced by interactions of cosmic rays
  in the atmosphere as one event per day in 5 kilotons of water.  
  
  Kamioka and IMB were the first large water Cherenkov detectors to see neutrinos 
  of extraterrestrial origin when they observed Supernova 1987A~\cite{Kamioka,IMB}.
  Kamioka had a fiducial volume of 4.5 kilotons and IMB was 8 kilotons.  Both detectors were 
  designed with the primary goal of searching for proton decay.  Because interactions of
  atmospheric neutrinos are the main background for proton decay, both detectors measured
  those neutrinos in as much detail as possible, including contained interactions
  of muon and electron neutrinos as well as $\nu_\mu$-induced muons, both through-going and
  stopping.  Hints of deviations from the expected ratio of muon to electron neutrinos
  showed up in both detectors, most notably in the energy and angular dependence of the
  ratio of electron-like to muon-like events reported by Kamiokande~\cite{Kam-88}.
  
  The second generation Super-K detector with 11,000 50 cm photomultipliers viewing
  50 kilotons of water was large enough to present convincing evidence of neutrino oscillations
  in the atmospheric neutrino sector in 1998~\cite{SK98}, soon after it began full operation.
  The spectrum of atmospheric neutrinos is now measured as a function of angle from
  ~100 MeV to tens of GeV for both $\nu_\mu$ and $\nu_e$, with oscillation parameters
  in the $\nu_\mu -\nu_\tau$ sector well determined~\cite{nuReview}.  In the meantime the 
  understanding of the deficit of solar $\nu_e$ as a consequence of neutrino
  oscillations was established conclusively with the heavy water
  Cherenkov experiment at the Sudbury Neutrino Observatory~\cite{SNO}.
  
  Starting in the 1970s the quest to instrument a much larger volume of water to make
  a deep underwater muon and neutrino detector (DUMAND) for high energy astrophysical
  neutrinos began.  Given the likely relation between sources of cosmic rays and high energy
  neutrinos of astrophysical origin, the expected level of the signal
  can be estimated~\cite{WB,GHS,Gaisser}.  Because of the small neutrino cross
  section, the largest possible volume is needed, and this
  inevitably leads to relatively coarse instrumentation and high energy threshold.
  For example, the ratio of photocathode area to volume in IceCube is only $\sim0.25$~cm$^2$/kT
  as compared to Super-K with $11000\times 2000\,{\rm cm}^2/50$~Kt, which is more than a
  million times bigger.  Correspondingly, the energy threshold for IceCube is 
  $> 100$~GeV or higher as compared to $< 10$~MeV for Super-K.
  Although DUMAND itself was realized only with the deployment of a
  single string for several days in 1987 from a ship~\cite{DUMAND1}, 
  the DUMAND effort in the seventies and eighties set the stage for high energy neutrino astronomy.
  The Baikal detector~\cite{Baikal} and the ANTARES detector~\cite{Antares}
  are the two large neutrino telescopes currently operating in water.
  
  One of the first papers~\cite{HLS}
  to discuss the possibility of using ice rather than water as the detector medium
  was presented in 1989 at a conference on prospects for
  astrophysics in Antarctica~\cite{volume}.  The meeting was hosted by Martin Pomerantz
  and the Bartol Research Institute at Delaware with support from the NSF Office of 
  Polar Programs.  Plans for AMANDA (Antarctica Muon and Neutrino Detector Array) 
  developed in the decade following this meeting.
  It is interesting to note that the Center for Astrophysical Research
  in Antarctica (CARA) for millimeter and submillimeter astronomy at the South Pole
  traces its origin to the same meeting.  Five years later the Martin A. Pomerantz Observatory
  (MAPO) was inaugurated at the South Pole to house the astronomy experiments and AMANDA.
  The South Pole Telescope and IceCube are both descendants of that era.
  
  When it was complete in 2000, AMANDA-II consisted of 19 strings with a total of 677 optical modules.
  Analog signals were sent to the surface over a mixture of copper and optical fiber
  cables to electronics modules in MAPO.  Data were recorded on tape and sent at the
  beginning of each Austral summer season for reconstruction and analysis of events
  in the North.  AMANDA ran in its full configuration starting in 2000.   
  Searches for neutrino sources with AMANDA alone cover the
  period from 2000 to 2006~\cite{AMANDA}.  The details of the 6595 AMANDA neutrino candidates
  from this period are available on the web~\cite{url}.  
  AMANDA continued to run as a sub-array of IceCube until it was shut off May 11, 2009.  String 18
  of AMANDA was equipped and used to test the digital technology for IceCube\cite{Str18}.
  \vfill\eject
  \section{Design, construction and operation of IceCube}
  
  Each optical module in IceCube is equipped with its own programmable data acquisition 
  board~\cite{DAQ} to digitize
  pulses from the photomultiplier (PMT)~\cite{PMT} 
  and to provide a time stamp for each event--hence the
  name {\it digital optical module} (DOM).  Times are keyed to a single GPS clock on
  the surface in such a way that the timing across the full array including IceTop
  is accurate to $<3$~ns.
  Each DOM has a local coincidence capability
  by which a condition can be applied to require a pulse above threshold in a
  nearest or next-to-nearest DOM as a condition for forwarding the full pulse to the surface.  
  This condition is called hard local coincidence (HLC).  In addition to the main board and
  the PMT, every DOM also contains a board with LED flashers for calibration.  
  The photomultiplier, main board, 
  calibration board and HV board are all housed in a 13" diameter glass sphere partially
  evacuated to 0.4 atmosphere.  A single penetrator connects the DOM cable to the electronics inside.
  
  The DOMs for IceCube were assembled and tested to strict standards by members of the
  IceCube Collaboration at three locations, one in the U.S., one in Germany and one in Sweden.
  They were tested again at the South Pole before deployment.  
  More than 99\% of the sensors survived installation and were 
  successfully commissioned.  Ninety-nine per cent of those commissioned 
  are fully functional after 16,000 DOM years.
  
  The standard IceCube string is a 2.5 km cable that carries the
  wires for 60 DOMs spaced at intervals of 17 m on the bottom kilometer of the cable
  (between 1450 and 2450 m below the surface).
  Breakouts on the main cable provide connectors for the DOM cables.
  Down-hole cables are arranged on a triangular grid with 125 m spacing on average.
  Surface cables connect to junction boxes near the top of each hole and carry the signal wires
  to the centrally located IceCube Lab.
  
  A key accomplishment that made IceCube possible was development of an enhanced hot 
  water drill (EHWD) system capable of drilling 60 cm diameter holes to a depth of 2.5 km
  efficiently and reliably.  The basic technique of drilling with hot water under high 
  pressure was used in AMANDA under the leadership of Bruce Koci.  The deepest
  AMANDA holes were drilled to 2 km.  It took three Antarctic seasons to work out
  the problems of drilling deeper holes for IceCube.  By the 4th season, the drilling and deployment 
  were able to proceed at the rate of 18-20 holes per season.
  
  \begin{table*}[thb]
\begin{center}
\begin{tabular}{lccccccc} \hline \hline
Season & 04/05 & 05/06  & 06/07 & 07/08 & 08/09 & 09/10 & 10/11 \\ \hline
Strings deployed & 1 & 8 & 13 & 18 & 19 & 20 & 7 \\ 
Total strings & 1 & 9 & 22 & 40 & 59 & 79 & 86 \\ \hline \hline
\end{tabular}
\end{center}
\end{table*}
  
    \begin{figure}
     \mbox{\epsfig{figure=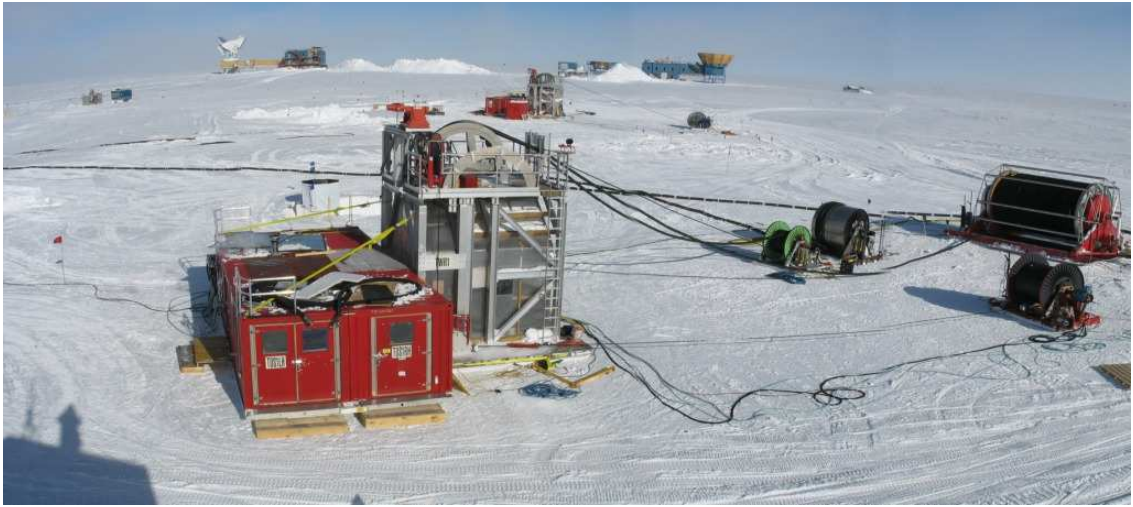,width=15.0cm}}
\caption{IceCube drilling and deployment operations.  The tower in the foreground is
drilling while the more distant tower is being use to deploy the string in the previously
drilled hole.  The South Pole telescope (left) and MAPO are visible on the horizon.}
\label{twotowers}
\end{figure}
  
  Several developments contributed to the success of the EHWD.  One was use of a single 2.5 km long
  hose for drilling.  Another was use of two drill towers in leapfrog fashion.
  The drill tower stands directly over the hole.  It carries the hose from the main hose reel as
  the drill head is lowered into the hole.  The same drill tower is used to carry the cable
  as it unwinds from its reel during deployment.  Sixty DOMs were staged for deployment in
  a milvan annexed to the tower.  Having two drill towers allowed drilling operations to begin
  at the next hole while deployment was in progress at the hole just drilled.  Another important
  development was the independent firn drill.  This device was used to melt a hole down to 50 meters,
  the depth at which the ice is sufficiently dense so that water pools and the more powerful
  hot water drill can be used.  The firn drill was independent of the drill tower so that holes
  could be prepared in advance, allowing main drilling to begin as soon as the
  tower was positioned.
  The drill was supplied with hot ($88^\circ$~C) water and high pressure (1000 psi) by a system
  of hot water heaters and high pressure pumps recycling water in a closed system that
  included three main components: 
  1) a large reservoir (``Rodwell") in the ice, 2) up to 300 m of insulated hose on the surface
  from the pumps and heaters
  to the drill site, and 3) the hole and a return water hose.  The photograph in Fig.~\ref{twotowers} 
  shows two towers in operation,
  one drilling and one deploying with the South Pole Telescope and MAPO in the background.
  
  Typical drill time was 30-40 hours per hole, and deployment took 10-12 hours from the
  beginning of the cable drop (longer when
  special devices were deployed).  For given conditions of temperature and pressure, the speed
  of the drill on the way down determined the initial diameter of the hole, which was also
  affected by the water flow and speed of the drill on its way up.  Drilling was planned
  to produce a hole with a minimum lifetime at full radius of 30 - 40 hours after removal of the drill,
  depending on whether dust logging or ancillary deployments were planned.
  The time for complete refreezing depends on depth, ranging from two weeks near the bottom of the
  hole where the ice temperature is $\approx -20^\circ$~C to several days around
  1.5 km where the temperature is $\approx -40^\circ$~C.  The properties of the
  refrozen ice are currently being studied with a camera that was deployed
  in the last hole.
  
  Deployment of IceTop proceeded in parallel with laying of the surface cables. Each IceTop station
  consists of two tanks separated from each other by 10 meters.  The two DOMs in each tank
  are connected to the surface junction box, which is between the tanks at a distance
  of 25 m from the corresponding deep string.  IceTop tanks were filled 
  with water from the drill system as soon as power was available through the surface cables.  Freezing the
  tanks was managed by an insulated freeze control unit to obtain clear ice free of air bubbles.
  With 2.5 tons of water in each tank, freezing time was 50-55 days.
  
  Surveyors determined the location of each IceTop DOM and the top of each hole to an accuracy 
  of a few centimeters.  Initial location of the deep DOMs was determined from the cable payout
  records and pressure sensor data.  
  In a second stage flashers were used to determine relative vertical offsets of strings.
  Reconstructed muon data and flasher data
  are used to monitor for shearing or other changes over time, which
  have so far not been observed.  Locations of deep DOMs are known to an accuracy of $<1$~meter.
  
  The central area of the array contains eight more densely instrumented strings with
  50 DOMs separated from each other by 7 meters
  at depths between 2100 and 2450 m, which is below the main dust layer in the glacier at
  the South Pole.  The other 10 DOMs are immediately above the dust layer.
  Most of the deep DOMs on the special strings contain PMTs with higher quantum efficiency.
  The fifteen strings in the center of IceCube, including the central
standard string and the ring of 6 surrounding standard strings, occupy a cylindrical volume 
in the clear ice below the dust layer with a 125 m radius and a 350 m height.
The goal of DeepCore is to define a fiducial
  volume with improved response at low energy and to use the surrounding IceCube as a veto
  to identify starting neutrino-induced events from all directions.  
  
  The trigger rate of the full IceCube is approximately 2.5 kHz with a seasonal variation of
  $\pm 8$\% correlated with temperature in the upper atmosphere where the high energy
  muons are produced.  This rate is dominated by cosmic-ray muons with sufficient energy to 
  penetrate to 1.5 km of ice, typically 500 GeV or higher at production.
  Events are processed continuously by a system of computers on the surface that currently
  looks for events with eight or more HLC hits within 5 $\mu$s in the standard strings or 
  $\ge 3$ hits in DeepCore.  A set of filters selects events for physics analysis, including
  upgoing muons, high-energy events, cascade-like events and air showers in IceTop.  Also
  events from the direction of the moon and the sun and events coincident with gamma-ray bursts
  are selected.  Approximately 5\% of all events are sent North
  by satellite.  In addition a short record for each event (direction, total charge, time)
  as well as monitoring data are included in the satellite transmission.
  \vfill\eject
  \section{Neutrino Astronomy}
  The basic analysis in IceCube is the search for point sources of extraterrestrial 
  neutrinos~\cite{PtSrc}.
  The techniques and the resulting sky map are presented in a separate paper at this
  meeting~\cite{Teresa}.  In searches for signals from specific sources, including galactic 
  supernova remnants, active galactic nuclei and gamma-ray bursts, the data on off-source regions
  can be used to establish the background.  Monte Carlo simulation is used primarily
  to optimize the search methods and to improve the sensitivity of the searches.  The 
  possibility of a correlation in space and/or time with a source identified electromagnetically
  would enhance the likelihood that the observed neutrinos are indeed of astrophysical origin.
  
    \begin{figure}
     \mbox{\epsfig{figure=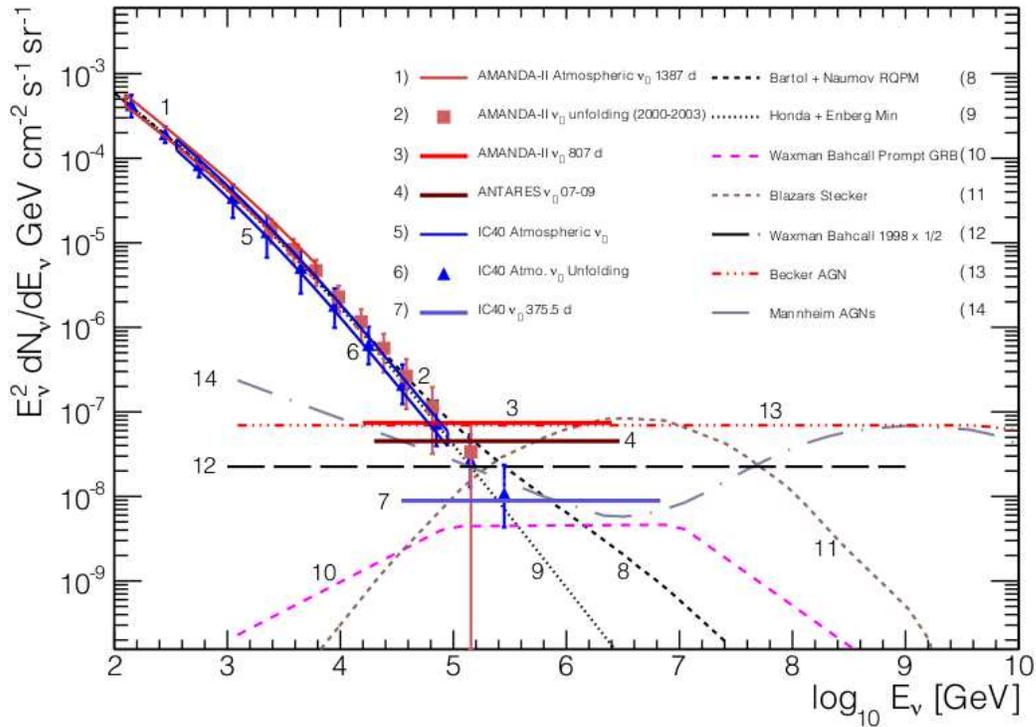,width=14.0cm}}
\caption{Summary of measurement, limits and models for diffuse fluxes of muon 
neutrinos\protect\cite{Sean}.  Also shown are measurements of the flux of atmospheric
neutrinos integrated over all angles.}
\label{Diffuse}
\end{figure}
  
  At the same time it is important also to
  search for an excess of astrophysical neutrinos from all
  directions above the background of atmospheric neutrinos.  
  Because the Universe is transparent to neutrinos, the total flux of neutrinos from all
  sources up to the Hubble radius is expected to be relatively large~\cite{Lipari1}.
  In general, astrophysical sources of neutrinos are expected to be distinguished from the 
  background of atmospheric neutrinos by their harder spectra.  In the IceCube
  energy range the atmospheric neutrinos are approaching asymptotic behavior in 
  which the spectrum becomes one power of energy steeper than the spectrum
  of cosmic rays incident on the atmosphere because the parent pions and
  kaons at high energy tend to interact before they decay.  Astrophysical neutrinos in general
  will be produced in a more diffuse environment in which all pions and kaons decay.
  In addition, if the astrophysical
  neutrinos are produced by cosmic rays interacting with radiation or gas in or 
  very close to their sources, then it might be expected that the neutrinos would
  have the same spectral index as the cosmic rays at their sources rather than the
  spectrum observed at Earth after propagation.  Differential spectral indices of
  $-2.0$ to $-2.4$ are generally expected for astrophysical neutrinos as compared to
  a differential index of $\sim -3.7$ for high energy atmospheric neutrinos.
  For extragalactic neutrinos a differential spectrum $\sim E^{-2}$ is often
  assumed, but this need not be the case.
   
  Figure~\ref{Diffuse} shows the current limit from IceCube~\cite{Sean} on a flux of
  astrophysical neutrinos from all directions assuming an $E^{-2}$ spectrum.
  The figure also shows measurements of atmospheric $\nu_\mu+\bar{\nu}_\mu$.
  References and explanations of the various model curves are given in
  the paper.\cite{Sean}
  It is clear from the figure that a good understanding of the background
  of atmospheric neutrinos around $100$~TeV and above is needed to
  see a component of astrophysical neutrinos, which would appear as
  a hardening of the measured neutrino spectrum above what is expected from
  atmospheric neutrinos.  
  
  An analytic approximation displays the key features
  of the atmospheric spectrum of muon neutrinos:
{\small  \begin{equation}
\phi_\nu(E_\nu) =  \phi_N(E_\nu)
  \times  \left\{{A_{\pi\nu}\over 1 + 
B_{\pi\nu}\cos(\theta)E_\nu / \epsilon_\pi}
+{A_{K\nu}\over 1+B_{K\nu}\cos(\theta)E_\nu / \epsilon_K}
 +{A_{{\rm ch}\,\nu}\over 1+B_{{\rm ch}\,\nu}\cos(\theta)E_\nu / \epsilon_{\rm ch}}\right\},
\label{angular}
\end{equation}}
\noindent
where $\phi_N(E_\nu)$ is the primary spectrum
of nucleons ($N$) evaluated at the energy of the neutrino~\cite{Gaisserbook,Lipari}
and $\theta$ is the zenith angle at the effective production height~\cite{Lipari}.
The three terms in brackets correspond to production from leptonic
and semi-leptonic decays of pions, kaons and charmed hadrons respectively.  
There is also a contribution from decay of muons not shown here that becomes
negligible for $E_\nu$ in the TeV range and above.  The most important
contribution to the intensity of TeV neutrinos is
$K^\pm\rightarrow\mu^\pm\,+\,\nu_\mu\,(\bar{\nu}_\mu)$, with a smaller
contribution from $\pi^\pm\rightarrow\mu^\pm\,+\,\nu_\mu\,(\bar{\nu}_\mu)$.
These channels constitute the ``conventional" neutrino spectrum.
As the neutrino energy increases above the critical energy, $\epsilon/\cos\theta$ 
($\epsilon_\pi\approx 110$~GeV, $\epsilon_K\approx 820$~GeV) the spectrum 
steepens, asymptotically by one power of energy.  The approach to asymptopia
occurs first near the vertical and at higher energy near the horizontal.
In the TeV range and above
the conventional spectrum becomes proportional to $\sec\theta$ and is peaked near the horizon.

The critical energy for charm is of order $10^7$~GeV, so in the present
energy region of interest neutrinos from charm decay have a spectrum that is isotropic and
reflects the primary spectrum without becoming steeper.  The contribution from
charm decay is small and uncertain.  Eventually, however, it is expected to become
the dominant contribution to atmospheric neutrinos at some energy because of its
harder spectrum.  Because the charm component is isotropic and harder than the conventional 
neutrino spectrum, it constitutes an important and uncertain background in
the search for a diffuse flux of neutrinos, which is also isotropic.

The search for extraterrestrial neutrinos in IC40~\cite{Sean} is done by measuring the
spectrum of upward-going muons and looking for an excess of events at high energy above what
is expected from atmospheric neutrinos.  For muon energy in the TeV range and above,
radiative processes become important and the muon energy loss per meter depends
linearly on its energy.  Simulations that take into account the properties
of the ice are used to relate the Cherenkov light emitted
along the track to the muon energy loss in the detector and hence to estimate
the energy of the muon as it passes through the detector

\begin{figure}
\mbox{\epsfig{figure=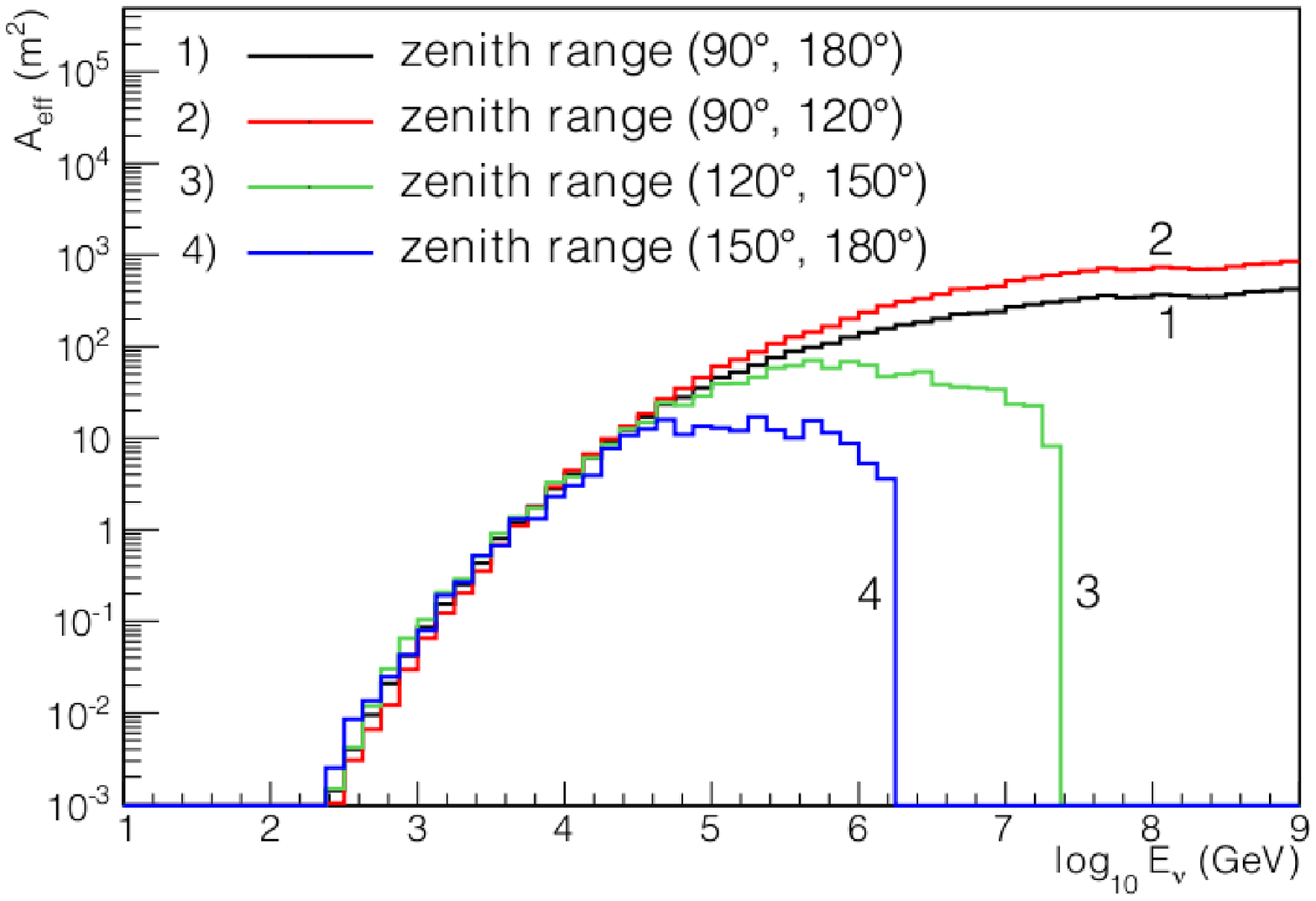,width=8.0cm}\epsfig{figure=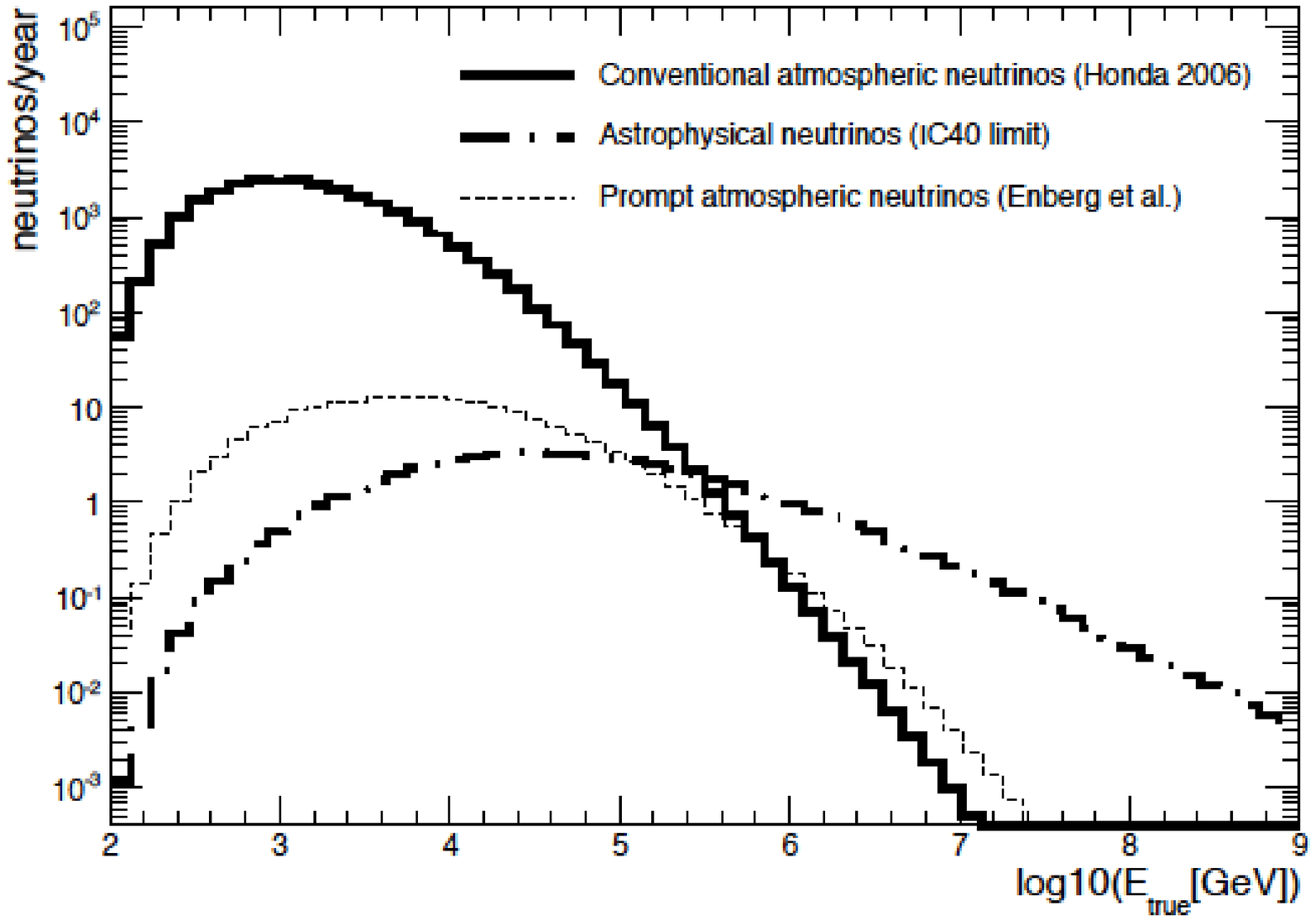,width=7.0cm}}
\caption{Left: $\nu_\mu$ effective area for IC40~\protect\cite{Sean}. 
Right: Response functions for conventional and prompt
atmosphere neutrinos and for an $E^{-2}$ spectrum in IC59.}
\label{Aeff40}
\end{figure}

The limit shown in Fig.~\ref{Diffuse} is obtained by a fitting procedure in which
the normalization of the conventional atmospheric flux, the normalization of
the prompt neutrino flux and the spectral index are allowed to vary
within an estimated range of uncertainty, for example, $\Delta\,\gamma < 0.03$
for the spectral index.  The shape of the conventional and prompt atmospheric
neutrinos are taken from Honda et al.~\cite{Honda07} and Enberg et al.~\cite{Enberg}
respectively.  Uncertainties in DOM efficiencies and ice properties were also
allowed to vary within estimated systematic uncertainties.  The normalization
of an astrophysical component with an assumed $E^{-2}$ spectrum was a
free parameter of the fit.  The data are entirely
consistent with the assumed conventional atmospheric neutrino spectrum, and the
best fit gave zero for the normalization of the prompt contribution as well as
for the astrophysical neutrinos.  The limit shown in Fig.~\ref{Diffuse} is
a 90\% confidence level upper limit assuming no contribution from prompt neutrinos.

Relating the flux of muons from below the horizon to the parent neutrino fluxes
requires a knowledge of the detector response folded with the assumed spectrum
of neutrinos.  The detector response is expressed as a neutrino effective area
as shown for IC40 in Fig.~\ref{Aeff40} (left).  $A_{\rm eff}(E_\nu,\theta)$ is a convolution of
the differential cross section for a neutrino with $E_\nu$ and zenith angle $\theta$ to produce a muon
with energy $E_\mu$ with the probability that the neutrino survives
propagation through the Earth to the interaction point and with the probability
that a muon produced with $E_\mu$ enters the detector with sufficient residual
energy to be reconstructed.  With this definition, the rate from a certain direction is  
$\int\phi_\nu(\theta)\times A_{\rm eff}\,{\rm d}E_\nu$.  There are four curves
in the plot of effective area, one for the average over all directions and three
others for specific bands of zenith angle below the horizon.  Half the
solid angle is in the range $90^\circ < \theta < 120^\circ$ where the flux
of atmospheric neutrinos is greatest.  Neutrinos in this zenith angle band
are produced in the atmosphere at Southern latitudes between $-30^\circ$ and $-90^\circ$.
Forty per cent of the horizontal zone is over Antarctica.  
Absorption in the Earth is minimal in the horizontal
band but becomes important for neutrinos from near the nadir already for $E_\nu > 30$~TeV.

An important aspect of interpreting the data is to account for the fact that
the range of neutrino energy that contributes to a given range of signals
in the detector depends on the assumed spectrum of the parent neutrinos.
This is illustrated in Fig.~\ref{Aeff40} (right) calculated for the 59-string version of
IceCube, which is currently being analyzed~\cite{Anne}.
The distributions of neutrino energy that give rise to the signal
in the detector are shown for three parent neutrino spectra: conventional atmospheric, prompt
atmospheric and $E^{-2}$ astrophysical neutrinos.
With the steep spectrum of conventional atmospheric
neutrinos, a muon of a given energy in the detector is likely to come from a neutrino
of relatively low energy that interacted in or near the detector.  In contrast, 
for an $E^{-2}$ spectrum, the same muon energy would more likely come from a neutrino
of much higher energy that may have interacted far from the detector.  	Analysis
of the response for IC40 shows that the range of neutrino energies from which 90\% of the events
comes is 300~GeV to 85~TeV for conventional atmospheric, 10 to 600~TeV for prompt and
35~TeV to 7~PeV for neutrinos with an $E^{-2}$ spectrum.  The latter range is plotted
for the upper limit in Fig.~\ref{Diffuse}.

The present analysis~\cite{Sean} has some limitations.  One is that the simulation
of the background of muons from atmospheric neutrinos is based on a calculation
of the spectrum of conventional neutrinos~\cite{Honda07} that extends only to 10~TeV.  
  A smooth power law extrapolation is
used to extend the tables up to the PeV range.  This does not
account for the steepening of the neutrino spectrum that must occur at some energy
as a consequence of the knee in the all-particle spectrum, which is around 
3~PeV total energy per particle.
Production of atmospheric neutrinos depends essentially on the spectrum of cosmic-ray
nucleons as a function of energy per nucleon, including both protons and nucleons in
helium and heavier primaries.  Thus the energy at which the neutrino spectrum steepens
depends also on composition of the primary cosmic rays.  In any case, the neutrino spectrum
is expected to steepen only for $E_\nu>100$~TeV, which is above the range of sensitivity
of the IC40 analysis for conventional atmospheric neutrinos.  
Another limitation is that the IC40 analysis 
integrates over all directions below the
horizon (but excluding a small angular region just below the horizon).  
The strong angular dependence of the conventional atmospheric neutrino
flux is a feature that can help distinguish this major background from
a small contribution of prompt neutrinos or extraterrestrial neutrinos.
Both the prompt neutrinos and the cosmic diffuse neutrinos would 
be isotropic (apart from absorption
in the Earth).  Both these limitations are being addressed in the current analysis
of IC59 data from 2009-2010~\cite{Anne}.  The new analysis will also use
an improved calculation of the neutrino effective area that includes a revised
treatment of the interaction geometry in the ice and rock below the detector.

As a by-product, the IC40 search for astrophysical neutrinos produces a measurement
of the atmospheric neutrino flux, shown by the blue lines (item \#5) in Fig.~\ref{Diffuse}.
This measurement is conservatively plotted in the 90\% sensitivity region for
conventional atmospheric neutrinos, which extends only to $85$~TeV.  Also shown is a
measurement of the atmospheric flux with AMANDA~\cite{AMANDAnu} as well as a
measurement with IC40~\cite{Warren}.  Both of these analyses use unfolding procedures
which provide extrapolations to somewhat higher energy.  The crossover
of the present diffuse upper limit and the extension of the measured spectrum
of atmospheric neutrinos is above 100 TeV, which sets the energy scale of interest
for future analyses.

Results from IceCube have so far emphasized muon neutrinos because of the good angular
resolution for the muon tracks, which is important in the search for point sources, 
and because of the large effective
area that follows from the fact that muons produced in charged current interactions
of neutrinos far outside the detector can be used.  IceCube is also designed to detect
cascades~\cite{cascades}, which can be produced by bremsstrahlung along a muon track, 
by charged current interactions of $\nu_e$ and $\nu_\tau$ 
and by neutral current interactions of any flavor neutrino.
Measurement of atmospheric electron neutrinos, for example,
is in principle more sensitive to the prompt atmospheric component than $\nu_\mu$ because the
spectrum of conventional atmospheric $\nu_e$ steepens at lower energy as muon decay
becomes unimportant.  Cascades in general are more sensitive
to astrophysical neutrinos because of the lower atmospheric background,
especially in the case of $\nu_\tau$.  The $\nu_\tau$ are rarely produced in
the atmosphere but are expected to be comparable to the other flavors
among extraterrestrial neutrinos because of oscillations and, in the PeV
range, to have the characteristic double bang signature~\cite{LearnedPakvasa}.  
The effective area for neutrino-induced cascades is smaller than for
$\nu_\mu$-induced neutrinos because the interactions cannot occur far outside the detector.
But it is easier to measure the neutrino energy, and atmospheric background is reduced.

\section{Implications}

For the first time the limits on high energy astrophysical neutrinos are below
the Waxman-Bahcall bound~\cite{WB}.  This level could be realized in models
in which ultra-high energy cosmic rays (UHECR) from extra-galactic sources accelerate
protons to high energy in an environment in which the accelerated protons are
magnetically confined in the acceleration region long enough to interact
with ambient radiation fields.  The photoproduction reaction
$p+\gamma\rightarrow p\pi^0$ would produce electromagnetic cascades initiated
by the $\gamma$-rays from $\pi^0$ decay.  The channel $p+\gamma\rightarrow n\pi^+$
would produce neutrons that leave the acceleration region and decay producing UHECR protons.
Neutrinos from decay of the charged pions would give a neutrino flux at a level
that is related to the contribution of the same sources to the intensity of UHECR.  
The Waxman-Bahcall limit gives an upper
limit for neutrinos produced by cosmic rays in sources transparent to nucleons.
The expected level of neutrinos could be lower (or higher) depending on how
the calculation is normalized to the high-energy end of the observed cosmic-ray spectrum,
but the present limit is beginning to challenge this class of models~\cite{Ahlers}.

A more specific model that relates neutrinos to UHECR is the gamma-ray burst model~\cite{GRB}.
IceCube is using a specific version of this model~\cite{Guetta} that provides a prediction of
the spectrum of neutrinos expected for each burst that depends on observed features
of the burst.  Limits based on data from two years of IceCube combining IC40 and IC59~\cite{D40}
are significantly below expectation ruling out much of the phase space for bulk Lorentz
factor of the GRB jet and its time scale~\cite{AGH}.

In general the limits IceCube is setting on neutrinos from extra-galactic sources
tend to disfavor models in which all sources of ultra-high energy (extra-galactic)
cosmic rays involve acceleration of protons inside compact sources.  An alternative
to such models, for example, would be acceleration by jets of AGN at their termination
shocks far from the active regions near the central black hole~\cite{Ferrari}.  In
this case, the accelerated UHECR would be a mixture of whatever ions are available
to be accelerated and neutrino producing interactions would be reduced.

\section{Outlook}

IceCube is a versatile detector that is breaking new ground in several areas
of astro-particle physics.  With an event rate approaching 100 billion per year, it is
possible to use the atmospheric muons to measure 
anisotropies in the cosmic-ray spectrum in the 10 to 500~TeV
range at the level of $10^{-4}$~\cite{anisotropy}.  Atmospheric muons have also
been used to calibrate the pointing and angular resolution of IceCube by
measuring the shadow of the moon\cite{moon}.
It will be possible to extend 
measurements of the atmospheric muon spectrum to approaching 1 PeV~\cite{Patrick}
and thus to look for prompt leptons in a way that is complementary to the
neutrino channel.  

Another analysis made possible by the size of the detector
and its high rate is the possibility to see details in the rates of TeV
muons in IceCube that reflect changes in the temperature profile of
the stratosphere above Antarctica~\cite{Tilav}.  The correlation between
seasonal variations and relative contributions of charm and kaons to the
neutrino flux has been noted~\cite{DG}.  If the prompt lepton flux can be measured
in the atmospheric muon channel, then its contribution to the background
in the search for a diffuse flux of astrophysical neutrinos can be accounted for.

Rates of hits above threshold in the DOMs are monitored continuously.  A galactic
supernova would show up as a sharp increase in counting rate due to the
light produced near the DOMs by many interactions of $\sim 10$~MeV neutrinos\cite{supernova}.  
Counting rates of DOMs in the
surface tanks can detect abrupt changes correlated with solar activity, and
particles from one solar flare have already been detected in this way~\cite{solar}.

Measurement of the flux of atmospheric neutrinos and its dependence on energy
and angle is of interest not only as background for astrophysical neutrinos,
but also because of the possibility of using the neutrino beam to probe
new physics.  The atmospheric neutrino analysis mentioned earlier~\cite{Warren}
also led to a new limit on a class of models in which violation of Lorentz invariance
produces an anomalous directional feature in the distribution of atmospheric neutrinos~\cite{Warren2}.
Another recent analysis used the angular dependence to set limits on
models of sterile neutrinos that produce new oscillation effects~\cite{Razzaque}.

A major effort of IceCube is the indirect search for dark matter.  Neutrinos from
WIMP annihilation in the Sun provide significant limits for WIMPs with large
spin-dependent interactions~\cite{solarWIMP}.  It is also possible to place
interesting limits on WIMP annihilation in the galactic halo~\cite{haloWIMP}.

Multi-messenger astronomy extends the concept of multi-wavelength astronomy
to neutrinos.  In addition to looking for neutrinos in coincidence
with photons from GRBs or flares of AGNs, there is also an active program to
send real-time alerts for followup in the optical or other wavelength bands.
For example, alerts are sent from IceCube to ROTSE and Palomar Transient Factory
whenever there are two or more neutrinos from within 3.5 degrees of each other
within 100 seconds\cite{Franckowiak}.  It is also possible to look for correlations
between neutrinos and gravitational waves\cite{Finley}.

IceCube, including its surface component, is a three-dimensional air shower array
with an aperture large enough to measure cosmic ray events up to one EeV
for events with trajectories that go through both IceTop and the deep
array of IceCube.  The ratio of surface signal to energy
deposition by the muon bundle in deep IceCube is sensitive
to primary composition~\cite{Feusels}.  Measuring the composition from the knee
region to high energy is of interest in connection with searching
for the transition from galactic to extragalactic cosmic rays.
Statistics in the EeV range can be improved by using
events with trajectories that go through the deep part of IceCube but pass
outside the perimeter of IceTop.  The direction of such events can be
reconstructed from the timing of hits in the deep detector alone
so that the hits in IceTop far from the shower core can be used to estimate
the shower size at the surface~\cite{COSPAR}.
Such events are also useful as an additional way to veto the background of
high-energy cosmic rays in the search for high-energy cosmogenic neutrinos.

Both neutrinos
and photons are expected to be produced in photo-production
processes as UHECR protons propagate through the cosmic background
radiation from sources at cosmic distances.  The neutrinos propagate freely,
but the photons cascade in extragalactic background light and show up
at lower energy where they contribute
to the diffuse gamma-ray background measured by Fermi~\cite{Fermi}.
A dedicated search for cosmogenic neutrinos with IC40~\cite{Aya} is
at the level where one event per year would be expected in the full
detector for a cosmogenic neutrino flux at the level of the upper
limit from Fermi on cosmogenic photons~\cite{Ahlers3}.  

Now that construction is complete and IceCube is running in its full configuration, 
the sensitivity of the search for astrophysical neutrinos will improve rapidly
as the total analyzed exposure increases.  One specific aspect that is
currently developing quickly is the ability to identify and measure cascades~\cite{cascades},
which will make possible a long planned feature of IceCube, namely, the ability
to distinguish neutrino flavors.  
With the complementary channels the full potential of IceCube can be realized.

\vspace{.1cm}
  
  \noindent
  {\bf ACKNOWLEDGMENT}: This work is supported in part by the U.S. National Science Foundation.


\begin{thebibliography}{99}
  \vspace{-.5cm}
  \bibitem{Reines} F. Reines, Ann. Rev. Nucl. Sci. 10 (1960) 1-26.
  \bibitem{Greisen} K. Greisen, Ann. Rev. Nucl. Sci. 10 (1960) 63-108.
  \bibitem{Markov} M.A. Markov, Proc. 1960 Annual International
  Conference on High Energy Physica at Rochester, ed. E.C.G.Sudarshan, J.H. Tinlot \& A.C. Melissinos
  (University of Rochester, 1960).
  \bibitem{Kamioka} K. Hirata et al. (Kamiokande-II Collaboration) Phys. Rev. Letters 58 (1987)
  1490-1493.
  \bibitem{IMB} R.M. Bionta et al. (IMB) Phys. Rev. Letters 58 (1987) 1494-1496.
  \bibitem{Kam-88} Y. Fukuda et al. (Kamiokande) Physics Lett. B335 (1994) 237-245.
  \bibitem{SK98} Y. Fukuda et al. (Superkamiokande Collaboration) Phys. Rev. Letters 81 (1998) 1562-1567.
  \bibitem{nuReview} L. Camilleri, E. Lisi \& J.F. Wilkerson, Ann. Revs. Nucl. Part. Sci. 58
  (2008) 343-369
  \bibitem{SNO} N. Jelley, A.B. McDonald \& R.G.H. Robertson, Ann. Revs. Nucl. Part. Sci. 59
  (2009) 431-465.
  \bibitem{WB} Eli Waxman \& John Bahcall, Phys. Rev. D59 (1998) 023002.
  \bibitem{GHS} T.K. Gaisser, Francis Halzen \& Todor Stanev, Physics Reports 258 (1995) 173-236.
  \bibitem{Gaisser} T.K. Gaisser, arXiv:astro-ph/9707283v1, 25 Jul 1997.
  \bibitem{DUMAND1} J. Babson et al. (DUMAND Collaboration), Phys. Rev. D42 (1990) 3613-3620.
  \bibitem{Baikal} V.A. Balkanov et al., (Baikal) Astropart. Phys. 12 (1999) 75-86.
  \bibitem{Antares} M. Ageron et al. (Antares Collaboration) arXiv:1104.1607v2, 13 June 2011.
  \bibitem{HLS} Francis Halzen, John Learned, Todor Stanev, in AIP Conf. Proc. {\bf 198} (1989) 39-51.
  \bibitem{volume} {\em Astrophsyics in Antarctia} AIP Conf. Proc. {\bf 198} 
  (ed. Dermott J. Mullan, Martin A. Pomerantz \& Todor Stanev, 1989).
  \bibitem{AMANDA} R. Abbasi et al. (IceCube Collaboration) Phys. Rev. D79 (2009) 062001.
  \bibitem{url} http://www.icecube.wisc.edu/science/data.
  \bibitem{Str18} M Ackermann et al. (AMANDA Collaboration) 
  Nucl. Inst. Methods A 556 (2006) 169-181.
  \bibitem{DAQ} R. Abbasi et al. (IceCube Collaboration) Nucl. Inst. 
  Methods A 601 (2009) 294-316. 
  \bibitem{PMT} R, Abbasi et al. (IceCube Collaboration) Nucl. Inst. 
  Methods A 618 (2010) 139-152. 
  \bibitem{PtSrc} R. Abbasi et al. (IceCube Collaboration) Astrophys. J. 732 (2011) 18.
  \bibitem{Teresa} Teresa Montaruli, this conference.
  \bibitem{Lipari1} Paolo Lipari, Phys. Rev. D78 (2008) 083011.
  \bibitem{Sean} R. Abbasi et al. (IceCube Collaboration) arXiv:1104.5187.
  \bibitem{Gaisserbook} T.K. Gaisser, {\em Cosmic Rays and Particle Physics} (Cambridge University 
  Press, 1990).
  \bibitem{Lipari} Paolo Lipari, Astropart. Phys. 1 (1993) 195-227.
  \bibitem{Honda07} M. Honda, Phys. Rev. D75 (2007) 043006.
  \bibitem{Enberg} R. Enberg et al., Phys. Rev. D78 (2008) 043005.
  \bibitem{Anne} A. Schukraft \& M. Walraff (for the IceCube Collaboration), Paper 0736, ICRC2011.
  \bibitem{AMANDAnu} R. Abbasi et al., Astropart. Phys. 34 (2010) 48-58.
  \bibitem{Warren} R. Abbasi et al. (IceCube Collaboration) Phys. Rev. D83 (2011) 012001.
   \bibitem{cascades} R. Abbasi et al. (IceCube Collaboration) arXiv:1101.1692.
   \bibitem{LearnedPakvasa} J.G. Learned \& S. Pakvasa, Astropart. Phys. 3 (1995) 267-274.
  \bibitem{Ahlers} M. Ahlers. L.A. Anchordoqui \& S. Sarkar, Phys. Rev. D79 (2009) 083009.
  \bibitem{GRB} E. Waxman \& J.N. Bahcall, Phys. Rev. Letters 78 (1997) 2292-2295 
  and Ap. J. 541 (2000) 707-711.
  \bibitem{Guetta} D. Guetta et al., Astropart. Phys. 20 (2004) 429-455.
  \bibitem{D40} R. Abbasi et al. (IceCube Collaboration) Phys. Rev. Letters 106 (2011) 141101.
  \bibitem{AGH} M. Ahlers, M.C. Gonzalez-Garcia \& F. Halzen, arXiv:1103.3421.
  \bibitem{Ferrari} Atillio Ferrari, Ann. Revs. Astron. Astrophys. 36 (1998) 539-598.
  \bibitem{anisotropy} R. Abbasi et al. (IceCube Collaboration) Astrophys. J. 718 (2010) L194-L198.  
  Also, arXiv:1105.2326.
  \bibitem{moon} D.J. Boersma, L. Gladstone \& A. Karle (for the IceCube Collaboration)
  Proc. 31st Int. Cosmic Ray Conf. (Lodz, 2009) arXiv:1002.4900.
  \bibitem{Patrick} Patrick Berghaus (for the IceCube Collaboration) Proc. 31st
  Int. Cosmic Ray Conf. (Lodz, Poland, 2009) arXiv:0909.0679.
  \bibitem{Tilav} S. Tilav et al. (for the IceCube Collaboration) Proc. 31st
  Int. Cosmic Ray Conf. (Lodz, Poland, 2009) arXiv:1001.0776.
  \bibitem{DG} T.K. Gaisser \& Paolo Desiati, Phys. Rev. Letters 105 (2010) 121102.
  \bibitem{supernova} Lutz K\"{o}pke (for the IceCube Collaboration) arXiv:1106.6225.
  \bibitem{solar} R. Abbasi et al. (IceCube Collaboration) Astrophys. J. 689 (2008) L65-L68.
  \bibitem{Warren2} R. Abbasi et al. (IceCube Collaboration) Phys. Rev. D82 (2010) 112003.
  \bibitem{Razzaque} Soebur Razzaque \& Yu.A. Smirnov, arXiv:1104.1390.
  \bibitem{solarWIMP} R. Abbasi et al. (IceCube Collaboration) Phys. Rev. D81 (2010) 057101.
  \bibitem{haloWIMP} R. Abbasi et al. (IceCube Collaboration) arXiv:1101.3349.
  \bibitem{Franckowiak} A. Franckowiak et al. (for the IceCube Collaboration)
  arXiv:0909.0631 (Proc. ICRC 2009).  Also paper 0445, ICRC2011.
  \bibitem{Finley} B. Baret et al., arXiv:1101.4669.
  \bibitem{Feusels} T. Feusels, J. Eisch \& C. Xu (for the IceCube Collaboration) Proc. 31st
  Int. Cosmic Ray Conf. (Lodz, Poland, 2009) arXiv:0912.4668.
  \bibitem{COSPAR} T.K. Gaisser (for the IceCube Collaboration), COSPAR 2010, arXiv:1107.1690.
  \bibitem{Fermi} A.A. Abdo et al., Phys. Rev. Lett. 104 (2010) 101101.
  \bibitem{Aya} R. Abbasi et al. (IceCube Collaboration) Phys. Rev. D83 (2011) 092003.
  \bibitem{Ahlers3} M. Ahlers et al., Astropart. Phys. 34 (2010) 106-115.
 
  \end{thebibliography}
  \end{document}